\documentclass[
prl,
nofootinbib,
twocolumn,
superscriptaddress,
showpacs
]{revtex4}

\usepackage{graphicx}
\usepackage{epsfig}
\usepackage{latexsym}
\usepackage{amsmath}
\usepackage{amssymb}
\usepackage{amsfonts}
\usepackage{amsxtra}
\usepackage{xcolor}
\usepackage{hyperref}
\def\etal{{\it et al.}}

\newcommand{\tr}{\text{Tr}}

\def\krto{ {\,\,\lower .8ex\hbox {$\longrightarrow \atop k \rightarrow 0$}\,\,}}

\def\bea{\begin{eqnarray} }
\def\beq{\begin{eqnarray} }
\def\eea{\end{eqnarray}}
\def\eeq{\end{eqnarray}}

\newcommand{\nn}{ \nonumber}

\newcommand{\chib}{{\bar \chi}}

\newcommand{\ben}{\begin{enumerate}}
\newcommand{\een}{\end{enumerate}}

\newcommand{\Bpsi}{{\bar \Psi}}

\newcommand{\vev}[1]{{\langle #1 \rangle}}
\newcommand{\Dslash}{{\not \hspace{-4pt} D} }

\begin{document} 

\title{Magnetic Catalysis in Graphene Effective Field Theory}

\author{Carleton DeTar}
\author{Christopher Winterowd}
\affiliation{Department of Physics and Astronomy \\ University of Utah, Salt Lake City, Utah, USA}
\author{Savvas Zafeiropoulos}
\affiliation{Institut f\"ur Theoretische Physik, Goethe-Universit\"at Frankfurt,
Max-von-Laue-Str.~1, 60438 Frankfurt am Main, Germany}
\begin{abstract}

We report on the first observation of magnetic catalysis at zero temperature in a fully nonperturbative simulation of the graphene effective field theory. Using lattice gauge theory, a nonperturbative analysis of the theory of strongly-interacting, massless, $(2+1)$-dimensional Dirac fermions in the presence of an external magnetic field is performed. We show that in the zero-temperature limit, a nonzero value for the chiral condensate is obtained which signals the spontaneous breaking of chiral symmetry. This result implies a nonzero value for the dynamical mass of the Dirac quasiparticle. This in turn has been posited to account for the quantum-Hall plateaus that are observed at large magnetic fields.

\end{abstract}


\maketitle





\section{Introduction}

Graphene, a novel material consisting of a single layer of Carbon atoms arranged in a hexagonal lattice, constitutes one of the most fascinating recent developments in condensed matter physics \cite{Novoselov}. This new field has seen tremendous growth due to the interest of both theorists and experimentalists in graphene's extraordinary electronic and mechanical properties \cite{CastroNeto, Goerbig}. Furthermore, there exists great hope for technological applications of graphene in electronic devices \cite{Segal}.  

From a theoretical point of view, graphene presents many puzzles and challenges. The electronic structure of graphene can be described by a valence and conductance band which touch at two inequivalent corners of the Brillouin zone. Near these so-called ``Dirac points", the dispersion is linear, and thus the quasiparticles can be described by two species of massless Dirac fermions. A low-energy effective field theory (EFT) can be constructed which describes these strongly interacting Dirac quasiparticles. 

In the presence of an external magnetic field, graphene exhibits a series of quantum-Hall plateaus that can be understood within a single-particle picture for Dirac fermions \cite{GusyninSharapov}. However, as the magnetic field increases ($B \geq 20~\text{T}$), new plateaus appear at filling factors $\nu = 0, \pm 1, \pm 4$ \cite{ZhangQHE,JiangQHE}. These plateaus are the result of the spontaneous breaking of the four-fold degeneracy of the Landau levels and the dynamical generation of a mass for the Dirac quasiparticle. Although Zeeman splitting appears to account for the plateau at $\nu=\pm4$, this ``magnetic catalysis" phenomenon is offered as an explanation for the other plateaus \cite{Yang, Kennett}.

Magnetic catalysis is a striking example of dynamical symmetry breaking whereby an external magnetic field induces fermion-antifermion pairing no matter how weak the attraction between the two (for a review see \cite{Shovkovy}). Originally studied in model relavistic field theories \cite{Miransky1, Miransky2}, it has also been predicted to appear in planar condensed matter systems such as graphene \cite{Khveshchenko, MiranskyGraphene1}. Previous approaches relied on perturbative, self-consistent techniques, and thus one would also like a fully nonperturbative approach.  Although previous lattice studies examined magnetic catalysis at finite temperature \cite{Polikarpov, Cosmai}, this is the first lattice study of the phenomenon at zero temperature. Preliminary results first appeared in \cite{DPF2015} and more details of our study will follow \cite{GrapheneLong}.

Even in the absence of interactions one can show that the chiral condensate, $\vev{\Bpsi \Psi}$, acquires a nonzero value due to the presence of the magnetic field. This supports the existence of spontaneous symmetry breaking (SSB). One can show that for Dirac fermions in $(2+1)$ dimensions in the presence of an external magnetic field \cite{Schwinger, DittrichGies}
\beq
\label{PBP2+1Final}
\lim_{m \to 0_+} \vev{\Bpsi \Psi} = -\frac{eB}{2\pi},
\eeq
where $e$ is the charge of the electron.
Although this result applies to the noninteracting case, one suspects that Coulomb attraction between particles and holes in graphene will strengthen the condensate and appropriately modify the result in (\ref{PBP2+1Final}). 

\section{Graphene EFT}
In the low-energy approximation, graphene is described by two species of massless Dirac fermions, arising from the points $K$ and $K'$, each having an additional degree of freedom coming from the spin of the original electron. These charged particles interact via the Coulomb interaction which is mediated by the scalar potential $A_0$. This approximation is reasonable due to the fact that the Fermi velocity of the quasiparticles satisfies $v_F/c \approx 1/300$. The Euclidean EFT describing these excitations reads
\beq
\label{ContinuumEFT}
\nn
 S_E &=& \int dt d^2 x \sum_{\sigma=1,2} \Bpsi_{\sigma} \Dslash[A_0] \Psi_{\sigma} \\ &+& \frac{(\epsilon + 1)}{4e^2} \int dt d^3 x ~ (\partial_i A_0)^2,
\eeq
where $\epsilon$ is the dielectric constant of the substrate on which the graphene monolayer resides, and the Dirac operator is given by 
\beq
\Dslash[A_0] = \gamma_0 \left( \partial_0 + i A_0 \right) + v_F \sum_{i=1,2} \gamma_i \partial_i.
\eeq
The four-component Dirac spinors are organized as follows:
\beq
\label{DiracSpinorBasis}
\Psi^{\top}_{\sigma} = \left( \psi_{K A \sigma}, \psi_{K B \sigma}, \psi_{K' B \sigma}, \psi_{K' A \sigma}\right),
\eeq
where $K, K'$ refer to the Dirac points,  $A, B$ refer to the sublattices, and $\sigma$ refers to the electron's spin projection. The graphene EFT possesses an internal $U(4)$ symmetry that is a consequence of the Dirac approximation. The generation of different mass terms allows for several patterns of SSB.

 To study the graphene EFT nonperturbatively, we will discretize the continuum action in (\ref{ContinuumEFT}) on a cubic lattice. The gauge sector is described by the ``noncompact" $U(1)$ lattice action \cite{Rothe}
\beq
\label{NCGaugeAction2}
S^{(NC)}_G = \xi \frac{\beta}{2} \sum_n \sum^{3}_{i=1} \left(\hat{A_0}(n) - \hat{A_0}(n+\hat{i})\right)^2,
\eeq
where $\xi \equiv a_s/a_t$ is known as the anisotropy parameter, which controls the ratio of the spatial lattice spacing to the temporal lattice spacing, $\hat{A_0}(n) = a_t A_0(n)$ is a dimensionless lattice field, and $\beta = (\epsilon + 1)/(2e^2)$. The Dirac action is discretized using staggered lattice fermions \cite{KogutSusskind}. In this formulation, the Dirac spinor structure of the action is diagonalized leaving
\beq
\label{GrapheneFermLatticeDimensionless}
\nn
S_F &=& \sum_{n} \bigg[ \frac{1}{2} \hat{\chib}_n \left(U_0(n)\hat{\chi}_{n + \hat{0}} - U^{\dagger}_0(n-\hat{0})\hat{\chi}_{n - \hat{0}}\right) \\ &+& 
\frac{v_F}{2\xi}\sum_{i=1,2} \eta_{i}(n) \hat{\chib}_n \left(\hat{\chi}_{n + \hat{i}} - \hat{\chi}_{n - \hat{i}}\right) \bigg],
\eeq
where $\hat{\chib}_n = a_s \chib_n$, $\hat{\chi}_n = a_s \chi_n$ are one-component, dimensionless Grassmann fields, $\hat{m} = a_t m$ is the dimensionless bare mass, and $\eta_1(n) = (-1)^{n_0}$, $\eta_2(n) = (-1)^{n_0 + n_1}$, are site-dependent phase factors. The interaction of the fermions with the scalar potential is introduced via the link variables $U_0(n) = e^{i e \hat{A_0}(n)}$. Staggered fermions preserve a remnant $U(1) \times U(1)_{\epsilon}$ symmetry left over from the continuum. 
The $U(1)$ symmetry, which corresponds to fermion number conservation, is given by
\beq
\label{U1Staggered}
\chi(x) \to \exp \left(i\alpha \right) \chi(x),~\chib(x) \to \chib(x) \exp \left(-i \alpha \right).
\eeq
The $U(1)_{\epsilon}$ symmetry, known as the ``even-odd'' symmetry, is given by 
\beq
\chi(x) \to \exp \left(i\beta \epsilon(x) \right) \chi(x), ~\chib(x) \to \chib(x) \exp \left(i \beta \epsilon(x) \right),
\eeq
where $\epsilon(x) \equiv \left( -1 \right)^{x_0 + x_1 + x_2}$. The $U(1)_{\epsilon}$ symmetry is a chiral symmetry, which is broken by the appearance of a mass term, $m\sum_n \hat{\chib}_n\hat{\chi}_n$. In practice, one explicitly introduces a mass term in order to regulate infrared singularities and to investigate SSB. After taking the infinite volume limit one then attempts to extrapolate to the chiral limit, $m \to 0$. In our simulations we use an improved version of the action in (\ref{GrapheneFermLatticeDimensionless}), known as the asqtad action \cite{Orginos}. The improvement removes leading discretization errors.

We now introduce a uniform magnetic field perpendicular to the sheet of graphene, described in the continuum by the vector potential $A_{\mu} = \delta_{\mu,2}Bx_1$. On the lattice this necessitates the introduction of static $U(1)$ link variables in the spatial directions
\beq
\label{ExtMagFieldLinks}
U_y(n) &=& e^{ia^2_seB n_x}, \\
U_x(n) &=& \left\{ \begin{array}{ll} 1 &, n_x \neq N_s-1 \\ 
                    e^{-ia^2_seBN_x n_y} &, n_x = N_s -1
                   \end{array} \right. ,
\eeq
where $N_s = N_x = N_y$ and $n_x, n_y = 0, 1, \dots , N_s-1$. The toroidal geometry (periodic boundary conditions in space) requires that $\Phi_B \equiv eB/(2\pi) = N_{B}/L^2_s$, where $L_s = N_s a_s$, is the lattice extent in the spatial direction and $N_B$ is an integer in the range $0 \leq N_B \leq N^2_s/4$ \cite{WieseAlHashimi}.

\section{Results and Discussion}
The main observable we use in determining SSB is the chiral condensate. In the chiral limit it serves as an order parameter distinguishing the semimetal phase, where it is zero, from the insulating phase, where it is nonzero. In the continuum, the chiral condensate is defined as follows:
\beq
\label{ChiralCondensateContinuum} \nn
\vev{\Bpsi \Psi} &=& \frac{1}{V}\frac{\partial \log Z}{\partial m}  \\ &=&  \frac{1}{V} \frac{1}{Z} \int \mathcal{D}A_0  \tr \left(\Dslash + m \right)^{-1}e^{-S^{\rm eff}_E[A_0]},
\eeq
where $S^{\rm eff}_E\left[A_0\right] = S_G\left[A_0\right] - \tr \log\left(\Dslash + m\right)$, and the partition function is given by
\beq
\label{PartitionFunction}
Z = \int \mathcal{D}A_0 e^{-S^{\rm eff}_E\left[A_0\right]}.
\eeq
On the lattice using staggered fermions, the equivalent expression is given by
\beq
\label{ChiralCondensateLattice}
\vev{ \chib \chi } = \frac{1}{V} \frac{1}{Z} \int \mathcal{D}A_0  \tr \left(\Dslash_{st} + m \right)^{-1}e^{-S^{\rm eff}_E[U_0]},
\eeq
where $\Dslash_{st}$ is the asqtad-improved stagggered Dirac operator and $m$ is the bare fermion mass. 

\begin{figure} 
\vspace{-1cm}
\includegraphics[height=8.5cm,width=8.5cm]{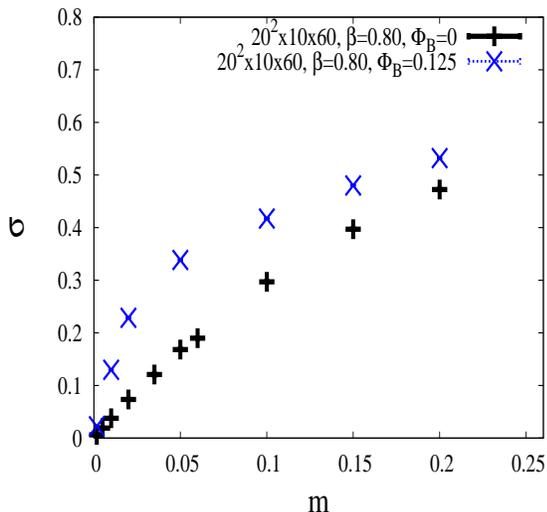} 
\vspace{-1.25cm}
\caption{The chiral condensate $\sigma \equiv \vev{\Bpsi \Psi}$ as a function of the bare fermion mass at zero field (black points) and at magnetic flux $\Phi_B=0.125$ (blue points), where flux is measured in units of $a^2_s$. We report the volumes in the form $N^2_s \times N_z \times N_{\tau}$. The error bars on each point are not visible on this scale.}
\label{fig:PBPComparison}
\end{figure}

We now outline our strategy for studying magnetic catalysis in the graphene EFT. At zero magnetic field, the graphene EFT is known to undergo a second-order transition as a function of the inverse coupling $\beta$ \cite{Drut}. Going to large values of $\beta$, one is firmly in the semimetal region, characterized by a vanishing chiral condensate. Fixing the coupling and turning on the external magnetic field, one would expect to see the condensate acquire a nonzero value. Our results for the chiral condensate as a function of bare mass at fixed $\beta=0.80$, for both zero and nonzero magnetic flux, are shown in Fig.~\ref{fig:PBPComparison}. One can see that for zero field, $\sigma \equiv \vev{\Bpsi \Psi}$ is approximately linear in the bare mass over the range of values shown. This is a clear indication that we are in the semimetal phase. At nonzero field, one can see that the condensate increases and exhibits strongly nonlinear behavior as a function of $m$. 

For both zero and nonzero magnetic flux, the condensate vanishes as the explicit symmetry breaking parameter is removed. When SSB occurs this may still happen because of the finite spatial extent of the box. We have explicitly checked this is not the case by calculating $\sigma$ as a function of $m$ for $N_s=8,20,30$. We find no significant depedence on $N_s$. One can explain this independence at nonzero magnetic field by noting that the magnetic length, $l_B \equiv \sqrt{\hbar c/eB}$, which characterizes the quasiparticle's cyclotron orbit, satisfies $1 < l_B < N_s$, in units of $a_s$. We have also checked the dependence of the condensate on $N_z$ and found that for $N_z \geq 10$, the finite volume correction is less than $2\%$. 

\begin{figure}
\vspace{-1cm}
 \includegraphics[height=8.5cm,width=8.5cm]{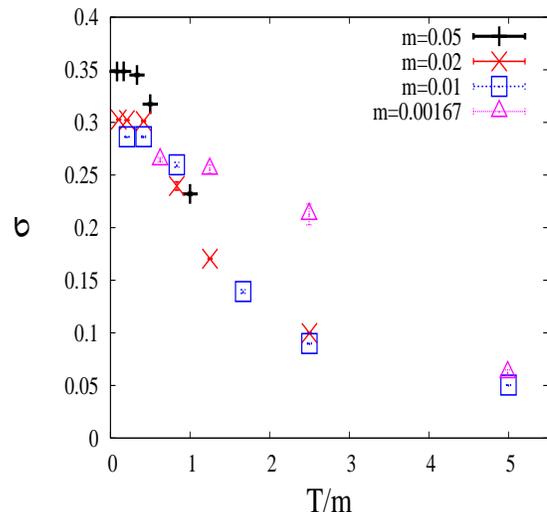} 
 \vspace{-1.25cm}
\caption{The chiral condensate $\sigma$ plotted as a function of $T/m$ for the ensembles with $\Phi_B=0.125$ and $N_s=8$.}
\label{PBPvsTdivM}
\end{figure}

The finite extent of the box in the Euclidean time direction, $N_{\tau}$, also plays a role. It is related to the temperature of the system, $T = 1/(N_{\tau}a_t)$. We vary $N_{\tau}$ to investigate the role of temperature  in restoring the symmetry. This is depicted in Fig.~\ref{PBPvsTdivM}. We find that when the dimensionless quantity $T/m$ is large the condensate tends to vanish while for $T/m < 1$ the condensate plateaus towards a finite value. To study magnetic catalysis in the ground state requires first taking the zero-temperature limit. Previous studies \cite{Polikarpov, Cosmai} were performed at finite-temperature and did not take this limit. 

\begin{figure} 
\vspace{-1cm}
  \includegraphics[height=8.5cm,width=8.5cm]{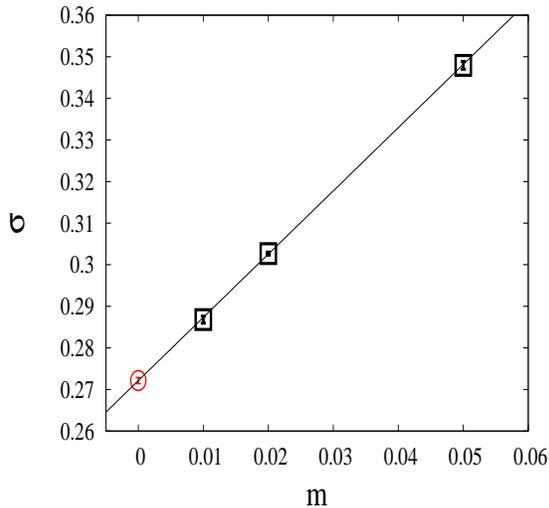} 
  \vspace{-1.25cm}
\caption{Chiral limit of the chiral condensate $\sigma$ using the $T=0$ extrapolated points ($\chi^2 \approx 0.6$). The errors on each point were determined from statistics as well as systematics i.e. the choice of model for the zero-temperature extrapolation.}
\label{fig:PBPzeroTChiral}
\end{figure}

\begin{figure} 
\vspace{-1cm}
 \includegraphics[height=8.5cm,width=8.5cm]{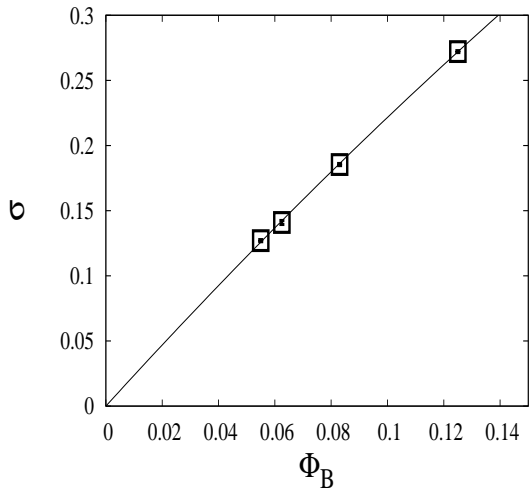} 
 \vspace{-1.25cm}
\caption{The chirally extrapolated, zero temperature values of the chiral condensate $\sigma$, plotted as a function of the magnetic flux,  $\Phi_B = eB/2\pi$.  The points at $\Phi_B=0.125$ and $\Phi_B=0.0625$ have a spatial size of $N_s=8$ while those at $\Phi_B=0.083$ and $\Phi_B=0.056$ have a spatial size of $N_s=12$. The errors on the points were obtained from the chiral extrapolations at $T=0$. We have fit the data to a quadratic constrainted to pass through the origin ($\chi^2/\text{d.o.f.} \approx 3.6/2$). } 
\label{fig:PBPzeroTChiralvseB}
\end{figure}

After taking the zero-temperature limit, one can then take the symmetry breaking parameter, in this case the bare mass $m$, to zero. In Fig.~\ref{fig:PBPzeroTChiral} we illustrate the chiral condensate as a function of the bare mass after taking the zero-temperature limit for magnetic flux $\Phi_B=0.125$. One can see that in the chiral limit a nonzero condensate is obtained. This procedure was then repeated for three other magnetic fluxes. The chirally extrapolated, zero-temperature results are displayed in Fig.~\ref{fig:PBPzeroTChiralvseB}. We thus have obtained a profile of the condensate as a function of the magnetic flux solidifying the scenario of magnetic catalysis in the graphene EFT.

\begin{figure}
\vspace{-1.2cm}
  \includegraphics[height=8.5cm,width=8.5cm]{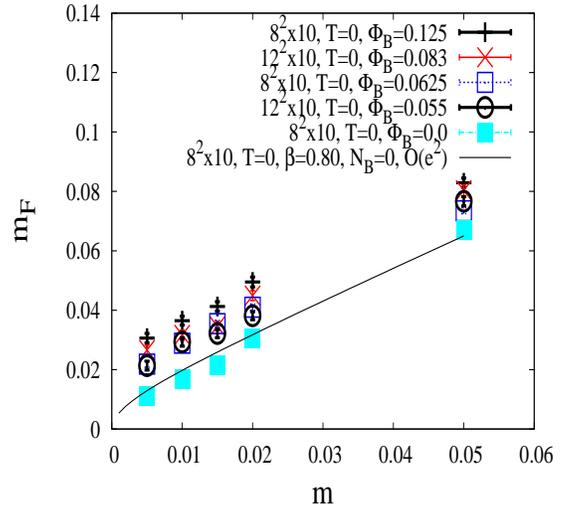} 
  \vspace{-1.25cm}
\caption{The fermion dynamical mass as a function of the bare fermion mass for all four nonzero magnetic fluxes and zero flux. The curve represents the pole of the fermion propagator ($\Phi_B=0$) calculated to $O(e^2)$ in lattice perturbation theory.}
\label{fig:MFvsmNonzeroB}
\end{figure}

SSB also has observable consequences for the fermion quasiparticle which acquires a dynamical mass, $m_F$. 	In order to study the dynamical mass, we calculate the quasiparticle propagator in the temporal direction at zero spatial momentum
\beq
\label{QuasiparticlePropagatorTemporal} \nn
G^{(\tau)}_F(\tau; \vec{p} = 0) &\equiv & \sum_{x,y} \vev{\chi(x,y,\tau) \chib(0,0,0)}, \\
&\sim & Ae^{-m_F \tau}, \quad \mbox{for large $\tau$},
\eeq
where $A$ is a $\tau$-independent constant.
The results for the dynamical mass are displayed in Fig.~\ref{fig:MFvsmNonzeroB}. One observes that at a given bare mass, the dynamical mass increases with the magnetic flux. Furthermore, the plot suggests that all four ensembles extrapolate to nonzero values in the chiral limit. Although one expects this to be the case for nonzero magnetic flux, at zero magnetic flux one expects the dynamical mass to vanish in the chiral limit. Perturbative arguments show that at zero magnetic field, the dynamical mass develops negative curvature near the origin and ultimately vanishes as the bare mass is removed \cite{GrapheneLong}.

\begin{figure}
\vspace{-1.2cm}
  \includegraphics[height=8.5cm,width=8.5cm]{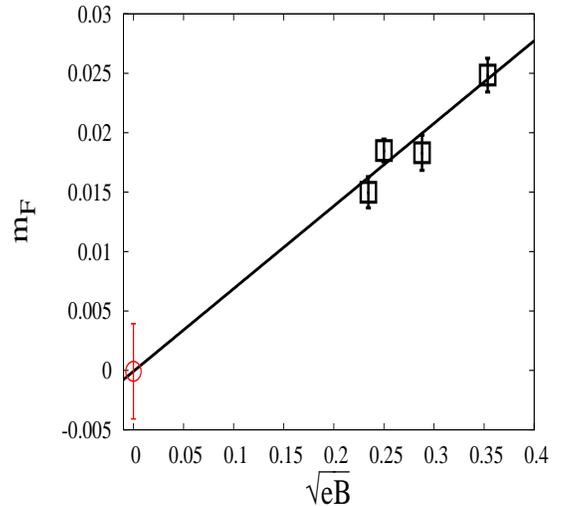} 
  \vspace{-1.25cm}
\caption{The zero-temperature, chirally extrapolated dynamical mass a function $\sqrt{eB}$. We have included a linear fit which gives an intercept which is consistent with zero ($\chi^2/\text{d.o.f.} \approx 3.7/2$). This is expected due to the fact that the dynamical mass vanishes in the chiral limit in the absence of an external magnetic field.}
\label{fig:MFvsSqrtB}
\end{figure}

Extrapolating the dynamical mass to the chiral limit for nonzero magnetic flux, we find a result which agrees with perturbative predictions for magnetic catalysis in $(2+1)$-dimensional field theories. Namely, we find that the chirally extrapolated quasiparticle mass, $m_F$, scales linearly with $\sqrt{eB}$. This is illustrated in Fig.~\ref{fig:MFvsSqrtB} giving further confirmation of magnetic catalysis in the graphene EFT.

\section{Conclusion}
Using nonperturbative methods for the study of the graphene EFT, we have shown the existence of spontaneous symmetry breaking at zero temperature due to an external magnetic field.
We have characterized the ground state of the system by performing a zero-temperature extrapolation of our observables followed by a chiral extrapolation. We have also characterized the dynamically generated mass of the Dirac quasiparticles and investigated its dependence on magnetic flux. In a future article we will present results for the time-reversal-odd Haldane condensate as well as the pseudoscalar Nambu-Goldstone boson that results from the spontaneous breaking of chiral symmetry \cite{GrapheneLong}.

\section{Acknowledgements} 
This work was in part based on a variant of the MILC collaboration's public lattice gauge theory code. See {\bf http://physics.utah.edu/$\sim$detar/milc.html}.
Calculations were performed at the Center for High Performance Computing at the University of Utah, Fermi National Accelerator Laboratory, and the LOEWE-CSC high-performance supercomputer of Johann Wolfgang Goethe-University Frankfurt am Main. We would like to thank HPC-Hessen, funded by the State Ministry of Higher Education, Research and the Arts, for programming advice. Numerical computations have used resources of CINES and GENCI-IDRIS as well as resources at the IN2P3 computing facility in Lyon. The authors would like to acknowledge discussions with Maksim Ulybyshev. SZ would like to acknowledge discussions with Wolfgang Unger. SZ would like to acknowledge the support of the Alexander von Humboldt foundation. CW and CD were supported by the US NSF grant PHY10-034278.



\end{document}